\documentstyle[11pt,paspconf]{article}
\input{psfig}

\newcommand{\rhom}{<\rho>}
\newcommand{\rhoc}{\rho_{\rm c}}
\newcommand{\Ho}{H_{\rm o}}
\newcommand{\dels}{\bar{\delta}}
\newcommand{\F}{{\mathcal{F}}}
\newcommand{\nuc}{\nu_{\rm c}}
\newcommand{\delc}{\delta_c}
\newcommand{\sigo}{\sigma_{\rm o}}
\newcommand{\Dg}{D_{\rm g}}
\newcommand{\sigeight}{\sigma_{8}}

\newcommand{\ng}{n_{\rm g}}
\newcommand{\fg}{f_{\rm g}}
\newcommand{\nel}{n_{\rm e}}
\newcommand{\Te}{T_{\rm e}}
\newcommand{\me}{m_{\rm e}}
\newcommand{\sigT}{\sigma_{\rm T}}
\newcommand{\Da}{D_{\rm a}}

\newcommand{\ata}{a_{ta}}
\newcommand{\xta}{x_{ta}}
\newcommand{\dmean}{\bar{\delta}}

\begin{document}

\title{Galaxy Clusters in Cosmology: \\
Cluster Abundance as a Probe of Structure Formation}
\author{J.G. Bartlett}
\affil{Observatoire de Strasbourg, 11 rue de l'Universit\'e,
	67000 Strasbourg, FRANCE}

\begin{abstract}
	In gaussian theories of structure formation, the galaxy
cluster abundance is an extremely sensitive probe 
of the density fluctuation power spectrum and of the
density parameter, $\Omega$.  We develop this theme by
deriving and studying in detail the mass function of 
collapsed objects and its relation to these quantities.
Application to current data yields constraints which
are degenerate between the amplitude of the perturbations
and the density parameter; we nevertheless obtain an
important limit on the present day {\em mass} perturbation 
amplitude as a function of $\Omega$ and can rule--out the 
`standard' cold dark matter (CDM) model. 
Future observations of the evolution of the cluster abundance 
will break the degeneracy and provide important constraints on
both the power spectrum and the density parameter, individually.
We focus primarily on X--ray clusters in the discussion, and
finish with a presentation of the promising new field of 
Sunyaev--Zel'dovich observations of cluster evolution.
\end{abstract}


\section{Introduction}

	Galaxy clusters are useful to cosmology  
because they may be studied as individual objects,
within the sea of the more general galaxy distribution,
with clearly definable dynamics.   Their characteristics
reveal much about the nature of the mechanism responsible
for the formation of large--scale structure
in the Universe, the central problem in modern cosmology.  
In this chapter, we shall consider in detail one 
cluster property -- their abundance and its evolution  --
and what it can tell us about our theoretical models.
I refer the reader to Sadat (these proceedings) for 
a discussion of some of the implications of other cluster 
attributes.
 
	Why should the cluster abundance be such a useful
tool for studying structure formation?  According to 
the favored scenario, formation by gravitational instability,
galaxies and galaxy clusters form where the density contrast,
$\delta$, (see Coles, these proceedings) is large enough that 
the surrounding matter may separate from the general expansion 
and collapse.  It should be no surprise, then, that
the abundance of collapsed objects depends on the amplitude
of the density perturbations.  These latter are modeled 
{\em statistically} and, on a given scale, they follow a probability
distribution, ${\rm Prob}(\delta)$ (e.g., a gaussian:
${\rm Prob}(\delta) = 1/\sqrt{2\pi\sigma^2}e^{-\delta^2/2\sigma^2}$).  
The amplitude of the perturbations on a scale, $R$, is defined as the 
variance, $\sigma(R)$, which is related to the {\em power spectrum}, 
$P(k)$, a quantity of fundamental importance (see below, and Coles,
these proceedings).  Now, the amplitude, $\sigma(R)$, 
appears to decrease with 
increasing scale, which implies that the density
contrast required to form a large object, such as a
galaxy cluster, becomes an increasingly rare 
event on the tail of the statistical distribution.
The present abundance of clusters is therefore expected
to be {\em extremely sensitive to small changes in} $P(k)$. 
In addition, the rate of cluster {\em evolution} is
essentially controlled by the density parameter, $\Omega$ 
(recall the linear--theory solutions
for density perturbations).  Thus, by studying the
cluster abundance and its evolution, we can constrain 
both $P(k)$ and $\Omega$.

	We begin with an introduction to the {\em mass function},
which provides a relation between the power spectrum
and the abundance of collapsed objects of a given mass.
After a brief look at the general problem of constructing 
the mass function, the Press--Schechter formula 
(Press \& Schechter 1974) is presented
as a sufficiently accurate approximation to 
the mass function observed in N--body simulations.
We then develop in detail its dependence 
on $P(k)$ and $\Omega$.  It is here that we derive 
the foundations of our approach.  In fact, 
application to real data is little more than 
finding trustworthy relations between the mass of a 
cluster and some more readily observable parameter, 
such as temperature.    

	In our application to data, we avoid optical 
cluster samples.  This omission is due, paradoxically,
to the difficulties of constructing well--defined
cluster catalogs as overdensities of galaxies; the
projection of galaxies along the line--of--sight
can result in cluster misidentifications and 
erroneous determination of cluster properties,
e.g., the velocity dispersion.  We instead
focus on X--ray observations.  The
empirical fact that the X--ray luminosity of a 
cluster scales as a large power of richness
(see Sadat, these proceedings, for a definition
of richness) eliminates projection effects
as a problem.  We give our attention primarily 
to the cluster X--ray temperature distribution
function.  We shall see why the X--ray temperature 
should be closely related to the mass, more
so than the X--ray luminosity, and how numerical
simulations indeed show a tight relation.  
We are then able to draw some important
conclusions from existing data 
concerning the power spectrum.  

	We conclude by studying the 
Sunyaev-Zel'dovich (SZ) effect.  In principle,
the cluster population may be studied by
this method in a manner completely analogous
to X--ray observations.  At the present 
time, however, this approach is much less advanced observationally, 
due to the rather demanding combination of radio 
telescope sensitivity, resolution and field--of--view 
requirements; it is, nevertheless, coming into its own right as 
an exciting new field.  We will examine the potential of 
this approach and also the advantages it offers, even in
comparison to X--ray observations.

\section{The Mass Function \label{jgb:sec:mf}}

	Our goal is to relate the number density of 
collapsed, virialized objects to the density perturbation
power spectrum.  These perturbations are characterized 
by their density contrast
\begin{equation}
\label{jgb:eq:dc}
\delta(\vec{x}) = \frac{\rho(\vec{x})-\rhom}{\rhom},
\end{equation}
where the mean cosmic density is denoted by $\rhom$ and
may be written as $\rhom = \Omega\rhoc = 
\Omega3\Ho^2/8\pi G$, introducing the density
parameter, $\Omega$ (we will also henceforth refer to the 
present value of the Hubble constant as $h\equiv \Ho/100\;$ km/s/Mpc).  
We usually work with a {\em
smoothed} version of the density field:
\begin{equation}
\label{jgb:eq:sdc}
\dels_R(\vec{x}) = \int d^3x^\prime\; \delta(\vec{x}^\prime)
	W_R(\vec{x}-\vec{x}^\prime).
\end{equation}
We will only consider a {\em top--hat} window function,
defined by
\begin{eqnarray}
\label{jgb:eq:tp}
\nonumber
W_R(\vec{x}) = &  W_R(x) = & \frac{3}{4\pi R^3} \quad x<R \\
\quad  & \quad & 0 \quad\quad {\rm otherwise}.
\end{eqnarray}
The {\em smoothing scale}, $R$, may be written in terms of 
the mass enclosed by the window: $M = (4\pi/3)R^3\rhom$.
Notice that the scale $R$ is written as a comoving 
coordinate, and so the enclosed mass is related to the
mean cosmic density, {\em not} the mean density of
a collapsed object.  We will more often speak of scales
in terms of mass, $M$, than in terms of linear scale, $R$.

	For completeness, I rewrite the expression for the
power spectrum, $P(k)$, in terms of the variance
of $\dels_R$:
\begin{equation}
\label{jgb:eq:psdef}
\sigma^2(M) \equiv <\dels^2_R(\vec{x})> = 
	\frac{1}{2\pi^2}\int dk\; k^2 P(k) |W_R(k)|^2.
\end{equation}
Strictly speaking, the average is over an ensemble of 
different realizations (different universes!) of the
density field at the point $\vec{x}$; pratically speaking,
the average is taken over space (i.e., positions $\vec{x}$).  
The Fourier transform of the top--hat window function used in Eq. 
(\ref{jgb:eq:psdef}) is  
\begin{equation}
\label{jgb:eq:tpk}
W_R(k) = \int d^3x\; e^{i\vec{k}\cdot\vec{x}}
	W_R(\vec{x}) = \frac{3}{(kR)^2} 
\left(\frac{{\rm sin}(kR)}{kR}-{\rm cos}(kR)\right) .
\end{equation}

\subsection{Derivation of the Press--Schechter Formula}

	In this section, we follow closely the notation
and presentation of Blanchard et al. (1992).
Let $\F(>M)$ be the {\em fraction of material} in collapsed objects
of mass $>M$.  The {\em mass function}, $n(M)$, giving the comoving
density of objects of mass $M$ per unit of mass, may then
be expressed as
\begin{equation}
\label{jgb:eq:mfdef}
n(M) dM \equiv \frac{\rhom}{M} \left|\frac{d}{dM} \F(>M)\right| dM.
\end{equation}
In hierarchical models, in which the perturbations strictly 
decrease with scale, and which include 
most of the currently favored models (e.g., cold dark matter
(CDM)--like models), we have the condition
\begin{equation}
\label{jgb:eq:mfnorm}
\int_0^\infty dM M n(M) = \rhom.
\end{equation}
This follows from the fact that each particle in the 
universe must find itself contained in a sphere whose 
probability for collapse approaches unity as the radius
decreases, because the variance becomes larger and
larger.   

	In many models, the density perturbations are gaussian
(important exceptions are defect models); this lead 
Press and Schechter (1974) to propose the following
ansatz:
\begin{equation}
\label{jgb:eq:psanz}
\F(>M) = \frac{1}{\sqrt{2\pi}}\int_{\nuc}^\infty d\nu e^{-\nu^2/2},
\end{equation}
where $\nu\equiv\dels/\sigma(M)$ and $\nuc$, or 
$\delc$, is some critical, {\em linear} 
density contrast required for an object to collapse.  There are
several important remarks concerning this
proposition: 
\begin{enumerate}
\item  It is assumed that the {\em linear} density contrast, 
	smoothed on a scale of mass $M$, is the only criterion 
	governing the formation of objects of this same mass, $M$;
\item  only points at the center of top--hat spheres are counted
	in the fraction $\F$;
\item  it is assumed that all objects with mass $>M$ are counted
	on a the scale $M$.  
\end{enumerate}
The first point is important because it means that we 
relate the abundance of {\em non--linear} regions 
to the {\em linear} theory density field, whose evolution
is simple to calculate.  Assumptions 
2 and 3 lead to the violation of 
condition \ref{jgb:eq:mfnorm}: $\int n(M) dM = 0.5 \rhom$ --
the ansatz misses some mass.  Press and Schechter originally
corrected this by simply multiplying Eq. (\ref{jgb:eq:psanz}) by 2.
More recently, Bond et al. (1991) showed clearly
that the missed mass is associated with objects of mass 
$>M$ which are not counted among the regions of mass $M$ 
capable of collapsing (remark 3).  This mass should also be closely 
related to particles within collapsed regions which do not find 
themselves at the center of collapsed spheres on the scale $M$ 
(remark 2).  Notice also that it is clear that the ansatz will
miss some mass, because as $M\rightarrow 0$, $\nuc\rightarrow 0$ and
not $-\infty$ to provide the necessary normalization; in other words,
the mass in underdense regions, $\nuc<0$, is poorly modeled at
the outset.  

\begin{figure}
\centerline{\psfig{file=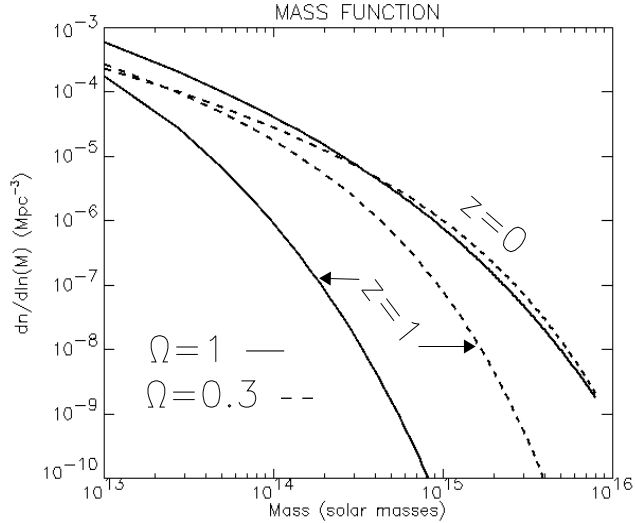,height=7.5cm,width=9cm}}
\caption{The mass function at two different redshifts.  The 
solid lines correspond to a critical universe with $n=-1.85$ and
a bias $b=1.65$.  The dashed lines are for an open universe with
$n=-1.4$ and $b=1.0$.  The parameter values are chosen to fit
the local X--ray temperature function (see below; Oukbir et al. 1996;
Oukbir \& Blanchard 1996).  In all cases, $h=1/2$.}
\label{jgb:fig:psm1}
\end{figure} 

	In any case, once corrected for the factor of 2, the
Press--Schechter (PS) mass function becomes
\begin{equation}
\label{jgb:eq:ps}
n(M,z) dM = \sqrt{\frac{2}{\pi}} \frac{\rhom}{M} \nuc(M,z) 
	\left| \frac{d{\rm ln}\sigma(M)}{d{\rm ln}M} \right| 
e^{-\nuc^2/2} \frac{dM}{M},
\end{equation}
where $\nuc(M,z)$ is as before, but now explicitly showing its
dependence on both mass and redshift (Problem 1).  For example, in a critical
universe ($\Omega=1$), linear perturbations grow as $1/(1+z)$, and
$\nuc = \delc(z)/\sigma(M,z) = \delc(z)(1+z)/\sigo(M)$, where 
$\sigo(M)$ is the present day power spectrum and $\delc(z)=1.68$,
a constant.  The value of this constant (1.68) can be found
by considering the collapse of a spherical overdensity (Peebles
1980).  More generally, we expect the critical density contrast
to depend on the redshift and the cosmological 
parameters, and we write $\delc(z;\Omega,\lambda)$,
where $\lambda$ is the cosmological constant.  
Even more importantly, the linear growth factor,
$\Dg(z)$, also depends on the cosmology:  $\Dg(z;\Omega,\lambda)$.
As already stated, for $\Omega=1$, $\Dg=1/(1+z)$.  With this 
notation we have
\begin{equation}
\label{jgb:eq:nu}
\nuc(M,z) = \frac{\delc(z;\Omega,\lambda)\Dg(0;\Omega,\lambda)}
	{\sigo(M)\Dg(z;\Omega,\lambda)}.
\end{equation}
We now see explicitly the dependence of the mass function on
the power spectrum and the cosmological parameters.

\subsection{Properties of the Mass Function}

	Let's explore in more detail the mass function by 
considering a simple power--law model for the power spectrum:
$P(k)\propto k^n$.  Via Eq. (\ref{jgb:eq:psdef}), this may also be 
written as 
\begin{eqnarray}
\label{jgb:eq:sigpl1}
\sigo(M) = \sigeight \left( \frac{M}{M_8} \right)^{-\alpha} \\
\label{jgb:eq:sigpl2}
\quad    = \frac{1}{b} \left( \frac{M}{M_8} \right)^{-\alpha},
\end{eqnarray} 
in which I introduce the mass contained in spheres of 
radius $8 h^{-1}$ Mpc as $M_8$.  The exponent $\alpha = (n+3)/6$.
The scale of $8 h^{-1}$ Mpc is often used as a reference for the
normalization of the power spectrum, because the variance
of galaxy counts in such spheres is about unity
(Davis \& Peebles 1983; Loveday et al. 1996).  This leads
to the notion of bias, and it is in the second line,
Eq. (\ref{jgb:eq:sigpl2}), that I introduce the 
{\em bias parameter} as $b=1/\sigeight$;  this quantifies any 
possible difference between the mass variance and the galaxy count
variance on this scale.

\begin{figure}
\centerline{\psfig{figure=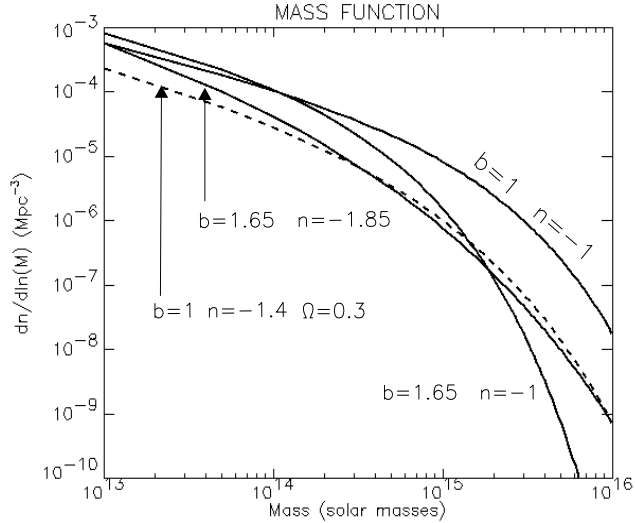,height=7.5cm,width=9cm}}
\caption{The influences of the power spectrum and density 
parameter on the PS mass function.  The power spectrum is
modeled by a power law -- $P(k)\propto k^n$ -- and 
normalized by the bias parameter $b=1/\sigeight$.  The solid
curves show the results for a critical universe, with the
labeled parameters.  The degeneracy over cluster scales 
between $P(k)$ and $\Omega$ is demonstrated by the dashed 
curve.  In all cases $h=1/2$.}
\label{jgb:fig:psm2}
\end{figure}

	With this power--law model, we can now write the 
PS mass function in a more compact and illustrative form:
\begin{equation}
\label{jgb:eq:pspl}
n(M,z) d{\rm ln}M = \sqrt{\frac{2}{\pi}} \frac{\rhom}{M_*} 
	\left( \frac{M}{M_*} \right)^{\alpha-1} \alpha
	e^{-0.5(M/M_*)^{2\alpha}} d{\rm ln}M.
\end{equation}
I have simplified things by defining a characteristic mass,
$M_*$, such that $\nuc(M,z) \equiv (M/M_*(z))^\alpha$.  
The characteristic mass depends on redshift; it increases with
time, or decreases with $z$.  Eq. (\ref{jgb:eq:pspl}) is plotted at 
two different redshifts in Fig. \ref{jgb:fig:psm1}.  The 
evolutionary trend is such that small objects merge to
form large ones, which progressively moves the ``knee'' towards large
scales and the low--mass end down.  This simple power--law
model is in fact quite useful and a not--to--bad approximation
over limited scales of more realistic power spectra.  
As we shall soon see, CDM--like models 
typically have values of $\sigeight\sim [0.5,1]$ 
and $n\sim [-2,-1]$ over cluster scales ($\ga 10^{14}$ solar masses).

	It is remarkable that this simple and appealing
PS formula provides an accurate description of the mass
function seen in numerical simulations, at least over
the interval of mass accessible to the simulations and
centered on the ``knee'' of the distribution.
At a given mass, the predicted  PS number of objects
deviates at worst from the numerical results by only
$\sim 25\%$.  This conclusion has been reached by 
by Eftathiou et al. (1988) for 
power-law spectra in a critical universe, by
Lacey and Cole (1994) using a larger simulation
of the same types of models, and more recently by 
Eke et al (1996), who considered CDM--like
power spectra in critical, flat and open geometries. 
It applies to the range of masses
of interest to us -- clusters with present--day virial 
temperatures $\sim 1-10\;$ keV.  Thus, it appears that, perhaps as
an happy accident, the PS formula provides 
a simple and accurate (to $25\%$ in number) representation of
the mass function over cluster scales for a variety of 
power spectra and cosmological models, within the context
of gaussian theories.

	Before we move on to apply what we have 
learned about the mass function to real
data, let me emphasize the essential elements of our aim.
We wish to deduce the mass function from cluster data and thereby
constrain $\Omega$, $n$ and $b$.
In Fig. \ref{jgb:fig:psm2} I plot the mass function at redshift
zero for several different power--law spectra in a critical 
and an open universe.  The set of solid lines shows the effects
of changing the power spectrum index ($n$) and normalization
($b$).  For the same normalization, ``flat'' -- or ``blue'' -- 
spectra (more negative values of $n$, represented
by $n=-1.85$) push mass up into larger objects than 
``steep'' -- or ``red'' -- spectra (represented by $n=-1$).
We also see quite clearly the sensitivity of the mass function
to the normalization.  The dashed line represents an open 
model ($\Omega=0.3$) with different 
values of $n$ and $b$ and which essentially reproduces one of the
critical models -- there is a degeneracy among 
$n$, $b$ and the underlying cosmological model.  

	Over cluster scales, the $n=-1$ power spectra 
mimic well the standard CDM spectrum ($\Omega=1$, $h=1/2$).
As we shall soon seen, this model does not correctly
reproduce the observed number of X-ray clusters, regardless
of the normalization.  The bluer spectrum ($n=-1.85$) 
fairs better, and is the reason for which I chose this
value for the figure (Oukbir et al. 1996).  
The open model parameters were chosen for the same reason
(Oukbir \& Blanchard 1996).

	The degeneracy between the power spectrum 
and the density parameter
prevents us from drawing any definitive conclusions 
from the mass function at a single redshift.  Fortunately,
the {\em evolution} of the mass function with redshift
breaks this degeneracy.  This is demonstrated by Fig. 
(\ref{jgb:fig:psm1}), where we see that degenerate
models evolve differently towards higher $z$.  The 
key point is that {\em in an open universe, one expects
more clusters at large $z$ than in an critical model}. 
The reason is that the linear growth factor, $\Dg$,
``freezes-out'' in an open model when the curvature
begins to drive the expansion; this is to say that
there is less growth at low redshift compared to a 
critical model.  
The consequence is that, normalized to the present day,
the cluster abundance evolves less rapidly towards
higher redshift in an open model.  I emphasize the
importance of this result, for it means that {\em the very
existence of high $z$ clusters carries important 
information about the density parameter, $\Omega$}.

\section{X--ray Clusters}

	In addition to galaxies, clusters consist of dark matter
(perhaps baryonic, perhaps not) and a diffuse, hot gas known as
the intra--cluster medium (ICM).  The trick is to get at the
mass of a cluster by observing these various components.
Traditionally, one measures the velocity distribution 
of a cluster's galaxies and applies the virial theorem (Sadat, these
proceedings).  As emphasized by several authors (e.g., Lucey 1983;
Frenk et al. 1990), the identification of cluster members is 
confused by the projection along the line--of--sight
of foreground and background galaxies, leading
to possible contamination of optical catalogs by false clusters;  
it seems clear, for example, that the 
Abell catalog is so contaminated (Dekel et al. 1989; Efstathiou et al.
1992).  In addition, the 
misidentification of cluster members inflates estimates of
cluster velocity dispersions, and hence virial mass
estimates.  The solution to these problems would seem
to be greater care in cluster selection, by using 
objective computer algorithms (rather than human 
inspection, as in the case of the Abell catalog), and
the acquisition of a large number of redshifts on each
cluster to better model its dynamics.  Such an approach 
becomes more observationally accessible thanks to the
ever increasing size of multi--object spectrographs.
However, because of these intricacies, we shall not consider
optical observations in the following.

	The ICM presents an alternative.  This gas, 
more or less in equilibrium in the cluster potential,
shines in bremsstrahlung emission at the virial temperature.
For a typical cluster, this temperature falls around a $few$ keV - 
clusters emit X--rays.  Detailed study
of cluster X--ray spectra demonstrates the thermal origin of 
the emission:  the continuum follows a thermal bremsstrahlung 
spectrum and the existence of the Iron emission lines
around 7 keV clenchs the conclusion.  

	Two important quantities characterize the X--ray
emission: the total (bolometric) luminosity and the
temperature.  The luminosity is an integral over the
cluster volume:
\begin{equation}
\label{jgb:eq:lx}
L_X \propto \int dV \ng^2 T^{1/2} \propto \fg M <\ng T^{1/2}>_{\rm p}.
\end{equation}
The second proportionality introduces the cluster gas fraction {\em by mass}
-- $\fg$ -- 
and the {\em particle averaged} quantity $<\ng T^{1/2}>_{\rm p}
\equiv (1/M)\int dV \ng (\ng T^{1/2})$.  The above equation gives us
a relation between the {\em observable} X--ray luminosity and
the cluster virial mass, $M$.  However, the application of 
this relation requires correct modeling of
$<\ng T^{1/2}>_{\rm p}$, and this depends most particularly
on the density of the gas.  This is to say
that the cluster luminosity is sensitive to the ICM's
spatial distribution, a direct consequence of the 
``$n$-square'' dependence of the bremsstrahlung process.
In terms of the King model (Sadat, these proceedings),
the X--ray luminosity depends on the core radius, $r_{\rm c}$, of 
the gas distribution.  The problem is that we really
have no understanding of the physics determining $r_{\rm c}$, 
and therefore modeling its evolution is difficult.
One may, nevertheless, approach the mass function by this direction
(Colafrancesco \& Vittorio 1994), but I prefer instead to move on to 
the temperature, which is much easier to deal with from 
a modeling perspective.  

	If we assume that the gas simply
shock heats on collapse to the virial temperature of the
gravitational potential, we would have
\begin{equation}
\label{jgb:eq:tx}
T \propto \frac{M}{R} \propto M^{2/3} (1+z),
\end{equation}
where $R$ is the virial radius of the cluster mass distribution,
i.e., the radius beyond which matter is infalling and has not
yet been dynamically incorporated into the cluster system.
The second proportionality follows from the expectation 
that $R\sim (M/\rho)^{1/3} \sim M^{1/3}(1+z)^{-1}$.  
In problem 2, I ask you to consider in detail the 
physics of collapse and to show that the {\em form} 
of this relation is general, but that the
exact value of the coefficient depends on
the mass profile of the collapsing cluster and on the
cosmological model.  Thus, in general $T={\rm coeff}
($cosmology, profile$) M^{2/3} (1+z)$.  As spelled out 
in the problem, the relation
follows from energy conservation only for the
{\em particle--weighted} temperature, $<T>_{\rm p} \equiv
(1/M)\int dV \ng T$.  Unfortunately, the measured X--ray
temperature of a cluster is {\em not} this quantity -
it is instead an emission--weighted version (this is not
true for the Sunyaev--Zel'dovich effect; more on 
this important point later!).
We may also note that what is truly observed is the
emission--weighted mean {\em electron} energy, and it 
has been pointed out that this may not be the
same as the proton temperature, at least in the outer
parts of clusters (Teyssier 1996). 

	Given these complications, one may nevertheless
consider the observed X-ray temperature as a ``good''
indicator of the mass.  This is most thoroughly demonstrated
by the simulations presented by Evrard et al. (1996).  In this work,
the authors demonstrate the existence at $z=0$ of a tight relation
between the {\em observable X--ray temperature} and 
$M^{2/3}$.  The simulations also supply the numerical value 
of the coefficient:
\begin{equation}
\label{jgb:eq:txcoeff}
T = (6.8 h^{2/3}\; {\rm keV})\; M^{2/3}.
\end{equation}
Simulated clusters in a variety of cosmological models follow
this same relation with only a 20\% scatter in mass at a 
given temperature.  The numerical value of the coefficient
turns out to be very close to the value one finds
for a simple spherical collapse with a flat profile 
(the simulation value is slightly lower, probably due to 
incomplete thermalization of the shocked gas).  The relation
is so good, in fact, that Evrard (1997) uses it as a
primary mass indicator for observed clusters when considering
the baryon fraction over an ensemble of clusters
(see also Sadat, these proceedings).

	This means that we can now calculate with some 
confidence the expected cluster X--ray temperature distribution 
function from the mass function -- the T--M relation was 
exactly what we needed.  Comparison with observations then
permits us to use our understanding of the mass function to
constrain the power spectrum and the density parameter.
Unfortunately, the temperature function has only been measured at $z=0$,
and we are currently unable to directly observe its evolution
with redshift.  It is true, of course, that clusters are detected 
in X--ray surveys out to redshifts now approaching unity, 
but the full spectrum needed to find $T$ requires many 
more photons than a mere 
detection -- hence, the present limitation to $z=0$.  
This will change with the next generation X--ray satellites 
(in particular, ASCA, AXAF and XMM).  As per our previous discussion, 
we expect to be limited by the degeneracy among the cosmological parameters 
and the power spectrum when fitting our predicted temperature 
distribution function to the $z=0$ observations.  

\begin{figure}
\centerline{\psfig{file=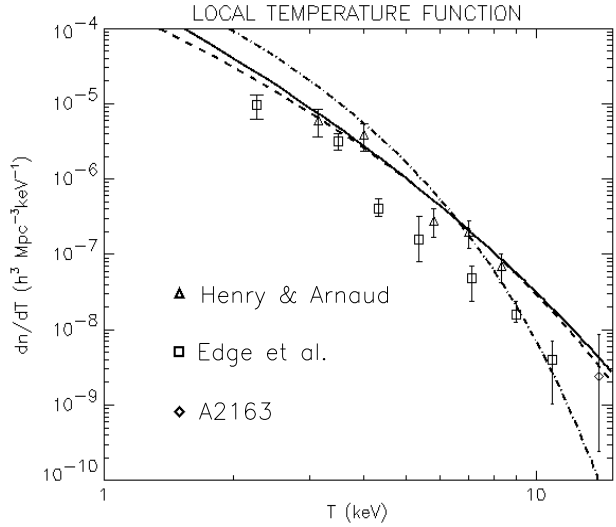,height=7.5cm,width=9cm}}
\caption{The present--epoch cluster X--ray temperature function
normalized to the data from Henry \& Arnaud (1991).
The solid line shows a critical model for a power--law
power spectrum with $n=-1.85$, $b=1.65$; the dashed
line is for an unbiased open model with $\Omega=0.3$, $n=-1.4$
(Oukbir et al. 1996; Oukbir \& Blanchard 1996).  A CDM spectrum 
normalized to $b=1.65$ produces the dot-dashed curve.
The diamond at 14 keV is an estimate by Oukbir \& Blanchard
based on A2163 (Arnaud et al. 1992).}
\label{jgb:fig:tfunc1}
\end{figure}

\begin{figure}
\centerline{\psfig{file=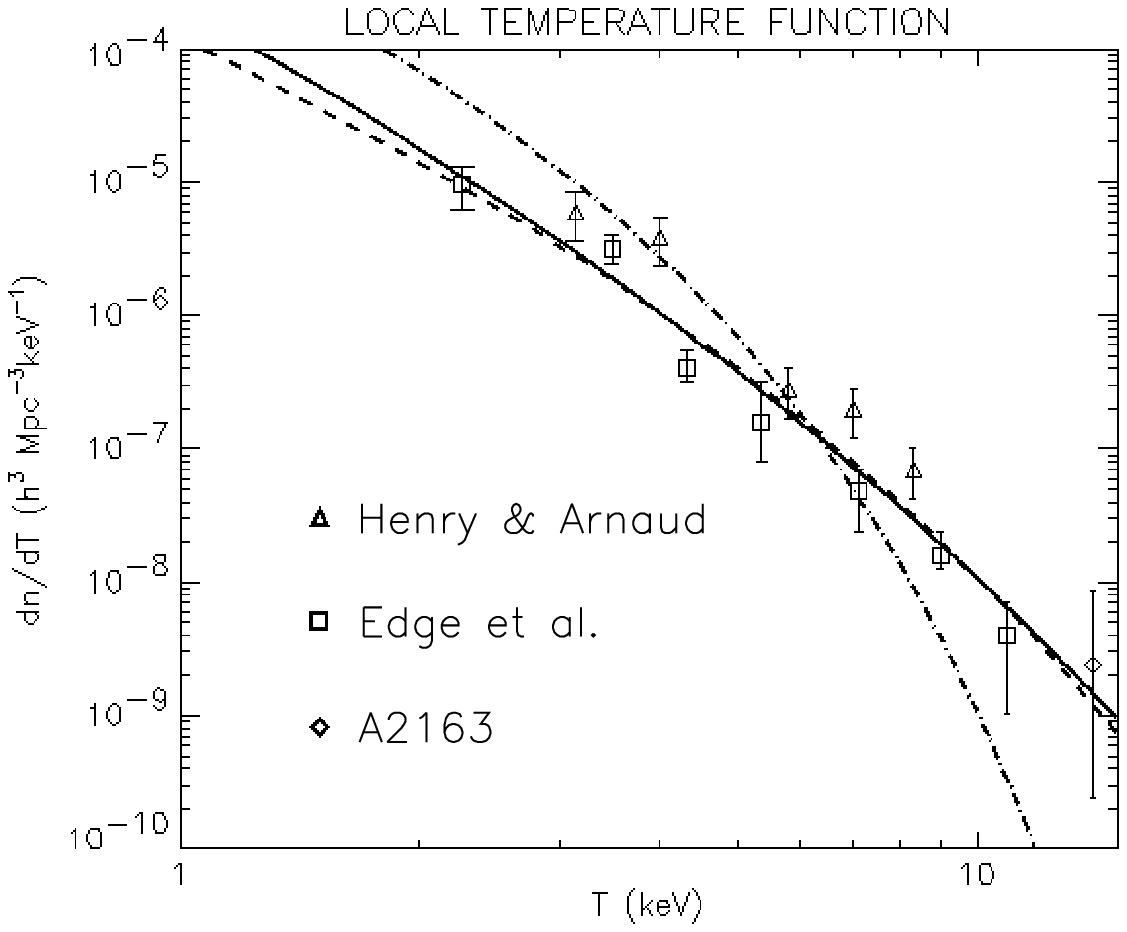,height=7.5cm,width=9cm}}
\caption{The present--epoch cluster X--ray temperature function
normalized to the data from Edge et al. (1990).  
The solid line shows a critical model for a power--law
power spectrum with $n-2.02$, $b=1.84$; the dashed
line is for an unbiased open model with $\Omega=0.2$, $n=-1.56$ 
(Oukbir et al. 1996; Oukbir \& Blanchard 1996).  A CDM spectrum 
normalized to $b=1.84$ produces the dot-dashed 
curve.  The diamond at 14 keV is an estimate by Oukbir \& Blanchard
based on A2163 (Arnaud et al. 1992).}
\label{jgb:fig:tfunc2}
\end{figure}

	In Figures \ref{jgb:fig:tfunc1} and \ref{jgb:fig:tfunc2}, 
I compare the observed local temperature function with some theoretical
predictions.  The data come from Henry \& Arnaud 
(1991) and Edge et al. (1990); 
it should be borne in 
mind that the two samples are {\em not} independent as they
share some of the same clusters.  The highest temperature
datum is an estimate by Oukbir \& Blanchard (1996) based
on the existence of A2163.  The theoretical calculations
have all been made by transforming the PS mass function, with
the given parameters, into a temperature function by using
the above T--M relation (for $h=1/2$).  Consider the power--law
models for $\Omega=1$ and $\Omega<1$.  We see that
in a critical universe clusters demand 
$\sigeight\sim 0.5 < 1$, i.e., {\em their abundance 
provides evidence for non--zero bias}.  However,
an open model with $\Omega<1$ and $\sigeight=1$ is
also consistent with the data, manifesting the
degeneracy we spoke about.  Thus, the cluster abundance can
be explained by either a critical, biased or open, unbiased scenario,
and this degeneracy can only be broken by observing
cluster evolution (more on this in a minute).  The
values of the best fit parameters for the critical
and open models were taken from Oukbir et al. 
(1996) and Oukbir \& Blanchard 
(1996).  Similar results for the
critical model, also based on power--law spectra,
were obtained by Henry \& Arnaud (1991).

	How about more realistic power--spectra? 
Several authors have worked with CDM--like power
spectra and used the local cluster abundance to
find the normalization, $\sigeight$ (Bond \& Meyers 1991;
Lilje 1992; Bahcall \& Cen 1993; Bartlett \& Silk
1993; White et al. 1993; Colafrancesco \& Vittorio 1994; Viana \&
Liddle 1996; Eke et al. 1996).  For a critical universe,
the normalization is insensitive to 
the exact shape of the power spectrum because
the scale of $8h^{-1}\;$ Mpc corresponds
to the mass of a rich cluster of galaxies:
$M_8 = (4\pi/3)\rhoc\Omega (8h^{-1}\;$Mpc$)^3 = 
5.9\times 10^{14}\Omega/h$ solar masses.  We expect, 
and it is the case, that the results should 
therefore be the same as found when using
power--law spectra.  Notice, though, that for smaller 
values of $\Omega$ the normalization scale moves out 
through the low end of the cluster mass range, 
introducing a correlation between the deduced 
value of $\sigeight$ and the shape of the power spectrum;
this is clearly demonstrated, for example, by the non--circular
contours in the $bias-n$ plane in the work of
Oukbir \& Blanchard (1996).  Thus, in principal,
the determination of $\sigeight$ by the cluster
abundance is model dependent in a low density universe --
it depends on the power--spectrum adopted.  

	As a summary of published results, which show a fair degree
of agreement, I quote Viana \& Liddle (1996):
\begin{eqnarray}
\label{jgb:eq:normresult1}
\sigeight \sim 0.60\Omega^{-0.59+0.16\Omega-0.06\Omega^2} 
	& \quad & \quad {\rm flat} \\
\label{jgb:eq:normresult2}
\sigeight \sim 0.60 \Omega^{-0.36-0.31\Omega+0.28\Omega^2}
	& \quad & \quad {\rm open} 
\end{eqnarray}
In their work, Viana \& Liddle fixed the shape of 
the power spectrum while varying the density parameter.
However, the CDM--like power spectrum depends on $\Omega$ 
and $h$, and we have just seen that the actual shape 
of the spectrum does influence the deduced value of $\sigeight$. 
For comparison, I quote a result obtained by taking this
into consideration (for $h=1/2$):
\begin{equation}
\label{jgb:eq:normresult3}
\sigeight = 1/(0.77 + \Omega - 0.04\Omega^{-0.35})
\end{equation}
(Blanchard \& Barbosa, private communication); the data
used where those of Henry \& Arnaud.

	The uncertainty in the final result is difficult to gage.  
The two data sets lead to normalizations 
differing by $\sim 10-20$\%.  Viana 
\& Liddle provide an error analysis based on their approach
(which does not take into account the detailed shape of 
the temperature function).  However, a thorough study of 
the various approaches with the goal of digging--out 
a more complete understanding of the errors has not
yet been performed.  As a final remark, I note
that Eke et al. (1996) claim that the high and low
ends of the observed temperature function are 
anticorrelated.

	The important conclusion is that the current
data (limited to $z=0$) on the cluster abundance
permit us to constrain the amplitude of the {\em mass} 
fluctuations at the present epoch.  It is clear, for 
example, that a critical model must be biased.  
 A second key result is that `standard' CDM does not work:
the shape of its power spectrum is too steep 
(too few large, hot clusters relative to the small ones)
and, in any case, a COBE normalization implies 
$\sigeight=1$, leading to far too many clusters at
any temperature.  There are, however, other models which 
do work.  Low-density models, either flat or open,
with $\Omega\sim 0.2-0.7$ and $h\sim 0.5-1$ satisfy 
{\em both} the COBE data {\em and} the cluster 
temperature distribution function; they also have 
power spectra similar to that seen in the galaxy 
distribution on large scales (Liddle et al. 1996a, 1996b).  
A critical model may be
made to work by changing its power spectrum.  
Tilting the primordial spectrum (Cen et al. 1992), adding some
hot dark matter (Davis et al. 1992; Schaefer \& Shafi 1992)
or lowering the Hubble constant (Bartlett et al. 1995)
are all ways of achieving the desired result.

	Can we say anything yet about evolution?
The deepest X--ray sample at the time of writing is the 
EMSS (Einstein Medium Sensitivity Survey; Gioia et al. 1990), which
contains $\sim 100$ clusters serendipitously detected in
Einstein pointings.  The clusters extend out to redshifts
approaching unity.  This catalog will soon be supplanted
by ROSAT samples based on the same principle - serendipitous
discovery in deep pointed observations.  Oukbir \& Blanchard
(1996) have modeled the
redshift distribution of the EMSS clusters by 
employing the locally observed luminosity--temperature 
relation.  This must be done because no direct
measurements of the temperature are yet available for 
these clusters.  It appears
that if there is no evolution in the L--T relation
with redshift, then a high--density model is 
favored; but with appropriate evolution, any
model can be made to fit the data.  For example,
both the critical, biased model and the unbiased,
open model with $\Omega=0.3$ shown in Figure \ref{jgb:fig:psm1}
can be made to work.  We must
await future observations with ASCA, AXAF and XMM,
which can directly measure the temperature 
of high redshift clusters, before any definite 
conclusion can be drawn; but the principle is
clear and the means will soon be available!

\section{Sunyaev-Zel'dovich Effect}
  
	The hot intracluster gas leads to another observational
consequence.  Via inverse Compton scattering, the electrons in 
the medium transfer energy to CMB photons passing through
the cluster and distort the CMB spectrum, an effect known as the 
{\em Sunyaev--Zel'dovich effect} (SZ) (Sunyaev \& Zel'dovich 1972).  
The effect is quantified
by the induced change in sky brightness, $i_\nu$, towards the cluster
as compared to the mean CMB intensity:
\begin{equation}
\label{jgb:eq:isz}
\delta i_\nu = y j_\nu(x),
\end{equation}
where the {\em Compton y--parameter} specifies the 
amplitude in terms of an integral along the line--of--sight  --
\begin{equation}
\label{jgb:eq:yparam}
y = \int dl \frac{k\Te}{\me c^2} \nel \sigT
\end{equation}
-- and the spectral dependence is given by
\begin{equation}
\label{jgb:eq:jnu}
j_\nu(x) = 2\frac{(kT_o)^3}{(h_pc)^2} \frac{x^4e^x}{(e^x-1)^2}
	\left[\frac{x}{{\mbox{\rm tanh}}(x/2)} - 4\right].
\end{equation}
In these expressions, $\Te$, $\nel$ and $\me$ refer to the
electron temperature, density and mass, respectively; 
$\sigT = 6.65\times10^{-25}\;$cm$^2$ is the Thompson
cross--section and $x\equiv h\nu/kT$ is the dimensionless
frequency of observation in terms of the {\em unperturbed
CMB} temperature $T=2.728\;$K (Fixen et al. 1996).  Note that $y$ is
dimensionless and $j_\nu$ carries units of sky brightness -
ergs/s/cm$^2$/steradian/Hz.  A convenient unit
is the Jansky (Jy), defined as $10^{-23}$ ergs/s/cm$^2$/Hz, so
that $j_\nu$ may be written in units of Jy/steradian.

\begin{figure}
\centerline{\psfig{file=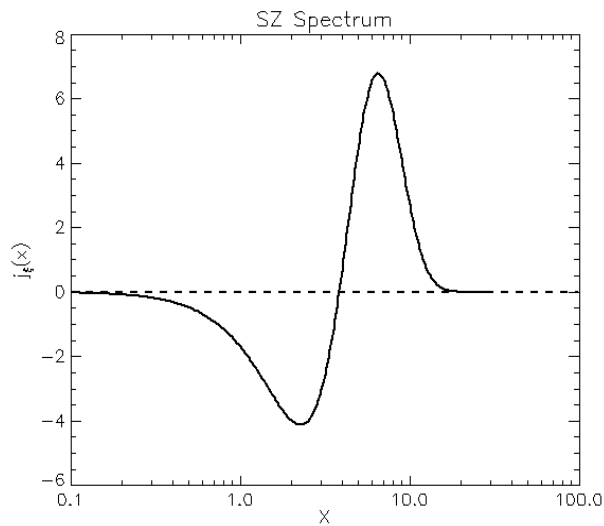,height=7.5cm,width=9cm}}
\caption{The SZ--induced change in sky brightness (relative to the 
unperturbed CMB) is plotted as a function of the dimensionless
frequency, $x$ (see text).  The units on the y--axis are meaningless.}
\label{jgb:fig:szspec}
\end{figure}

	I plot the spectral function ($j_\nu$) in 
Figure \ref{jgb:fig:szspec}; the shape is unique, unlike
any other astrophysical source.  This is the result
of the fact that, while increasing the mean photon 
energy, inverse Compton scattering conserves the 
number of photons.  In essence, the photons just diffuse 
upwards in energy, overpopulating the Wein region at the 
expense of the Rayleigh--Jeans.  Thus, a cluster appears
as a decrement in the sky brightness (a `hole') at
wavelengths longer than $\lambda=1.38\;$mm and as 
a source of excess brightness at shorter wavelengths.
I ask you in problem 3 to explore the physics of 
inverse Compton scattering and to derive Equation
(\ref{jgb:eq:jnu}).

	To give you a feeling for the order of magnitude
in rich clusters, the effect, which may be described in 
the Rayleigh--Jeans by a temperature perturbation towards
the cluster, is typically $\delta T/T\sim 
10^{-5}-10^{-4}$.  Incidentally, its detection proves
the cosmological origin of thenecessary normalization; in other words,
the mass in underdense regions, $\nuc<0$, is poorly modeled at
the outset.   CMB; the fact that this
is little doubted in the standard Big Bang model does
not diminish the importance of this result - such
proofs are always welcome.  

	Consider now the integrated effect of a cluster,
or its flux density, $S_\nu$:
\begin{equation}
\label{jgb:eq:Snu}
S_\nu(x,M,z) = j_\nu(x) \Da^{-2}(z) \int dV \frac{k\Te(M,z)}{\me c^2} 
	\nel(M,z) \sigT.
\end {equation}
The integral extends over the cluster volume and the angular distance 
$\Da(z)=2c\Ho^{-1}[\Omega z + (\Omega-2)(\sqrt{1+\Omega z} -1)]
/\Omega^2(1+z)^2$ (for $\lambda=0$).  The extremely important point here is 
that the {\em integrated effect of a cluster, its flux density, 
depends only on the quantity of gas at temperature $\Te$},
and {\em not on the spatial distribution of the gas}.
This is in contrast to the X-ray flux.  In fact, we may
write $S_\nu \propto \fg M <\Te>_p$, where the 
temperature is, again in contrast to X--ray observations,
the {\em mean, particle--weighted electron energy}.  Recall that this is the
quantity most clearly and directly related to the mass by 
energy conservation during collapse.  We may hope, therefore,
that the temperature--mass relation used here is not
too grossly affected by any non-isothermality of the ICM;
in any case, it should be much less so than the 
X--ray temperature.

\begin{figure}
\centerline{\psfig{file=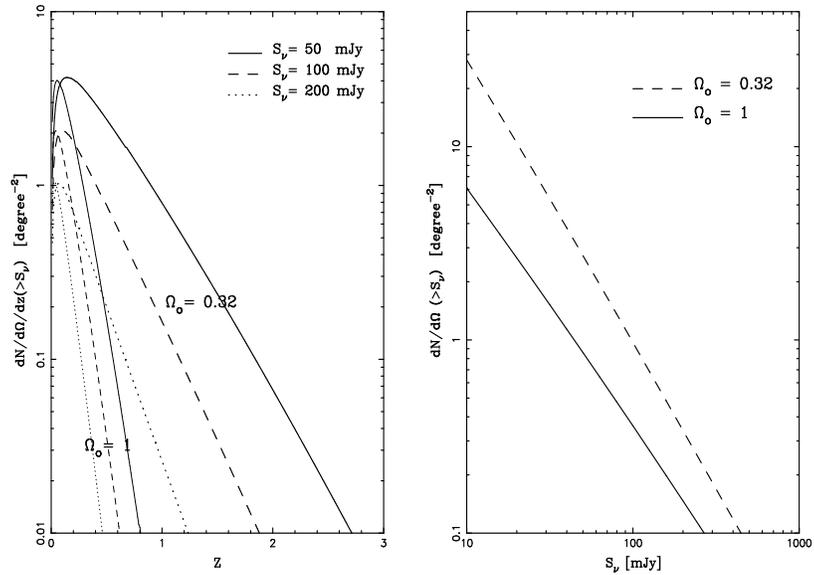,height=12cm,width=7.5cm}}
\caption{SZ redshift distribution and source counts.  On the
left, the thin lines trace the redshift distribution
for a critical model, at a given flux density, while the
thick lines represent the same for a degenerate open model.
The integrated counts for the two models are shown on the
right.  In all cases, $h=1/2$.  See Barbosa et al. (1996).}
\label{jgb:fig:szcounts}
\end{figure}

	Since the flux density is such a simple function
of the total mass and redshift of a cluster, we may,
as before, transform the mass function into an SZ 
distribution function at any redshift to calculate,
for example, the mean distortion produced by the
entire cluster population and the cluster number counts
and redshift distribution (Korolyov et al. 1986;
Markevitch et al. 1994; Bartlett \& Silk 1994; Barbosa et al. 1996).
Figure \ref{jgb:fig:szcounts} reproduces a plot
from Barbosa et al. (1996).  Actually 
testing these predictions is difficult, for it demands
a radio telescope of high resolution, about an arcmin,
capable of surveying large quantities of sky; one of 
the important goals for ESA's Planck Surveyor mission 
will be to measure the SZ cluster counts.  The
relatively simple modeling required to describe the
effect and the ability to observe clusters out to
arbitrarily large redshifts means that this 
method should prove very powerful in constraining
$\Omega$ and the history of the ICM.  

	I refer
the reader to Barbosa et al. for a discussion of
the implications for the mean $y$ distortion 
(see also Cavaliere et al. 1991; Markevitch et al. 1991);
it would appear that for low--density cosmologies,
the predicted distortion approaches the current
limit imposed by the FIRAS spectrum.  Finally, 
discussion of the fluctuations induced in the CMB
by unresolved clusters may be found in Shaeffer \& Silk 1988,
Cole \& Kaiser 1989, Bond \& Meyers 1991,
Markevitch et al. 1992, Bartlett \& Silk 1994, Ceballos \&
Barcons 1994, Colafrancesco et al. 1994.

\section{Conclusions} 

	As we have seen, galaxy clusters do indeed provide
one of the most important probes of the density perturbation
power spectrum and of $\Omega$.  The reason for this is clear:
Clusters find themselves on the tail of the rapidly falling
mass function expected in gaussian theories of structure
formation.  One may apply the same approach to non--gaussian
models, but only after recalculating the mass function.
The texture scenario is an example of a physically motivated
non--gaussian model, and an analysis of galaxy clusters
in this context may be found in Bartlett et al. 
(1993).  

	Many authors have used current observations of X--ray
clusters to constrain various popular theories of structure 
formation.  The results appear to all be in agreement: 
a critical universe requires a non--zero bias to 
avoid over--producing the number of X--ray clusters.
In particular, `standard' CDM does not work, its most 
the most severe shortcoming being the high normalization 
required by the COBE data; one could imagine that
gravity waves produce some of the COBE signal, thereby
lowering the implied CDM amplitude, but even in this case
it appears that the model produces a temperature distribution
function which is steeper than the data.  Quantitative results for the
amplitude, $\sigeight$, are given in Eqs. (\ref{jgb:eq:normresult1}),
(\ref{jgb:eq:normresult2}) and (\ref{jgb:eq:normresult3}).
The important point of this lecture should be that this 
method provides access to the
{\em mass} fluctuation amplitude, the fundamental 
theoretical quantity which is observationally rather elusive.

	The present data are limited to the local Universe, $z=0$,
which prevents us from separating the influences of the power
spectrum from the density parameter.  With the ability 
to measure cluster X--ray temperatures at large redshift,
future space missions, such as AXAF and XMM, will provide 
the means necessary to individually constrain the 
power spectrum and $\Omega$.  I would like to emphasize the
fact that clusters offer us a very useful probe of 
$\Omega$; this should be the second fundamental point of this
lecture.  

	We have not discussed models with non--zero $\lambda$, 
but the various given litterature references cover this topic.
In general, a flat, low--density model evolves with a rate 
somewhere between a critical model and the corresponding  
open model with the same density.  Such a scenario 
does not present any particular problems from the point
of view of cluster modeling.

	Statistical cluster studies via the SZ effect 
are just beginning to become an observational reality.
This line of research should be quite analogous to 
X--ray clusters studies; we thus have an independent
way to study cluster evolution.  In fact, the SZ
effect will permit us to see clusters to larger redshifts
than X--ray observations - provided they are out there, a
question whose answer has fundamental consequences.
In addition, the SZ effect is easier to model than
the X--ray emission of a cluster, an important
advantage.  By providing the first all--sky survey of 
clusters detected by their SZ effect, ESA's Planck
Surveyor (ex. COBRAS/SAMBA) seems to guarantee
an exciting future for this new field of cluster 
observations.

\section{Suggested Problems}


\subsection{The Mass Function}

	The mass function of collapsed objects is one of the
fundamental quantities of any theory of structure formation.
Following the notation of Blanchard et al. (1992), 
we write this function formally as
\begin{displaymath}
n(M)dM = -\frac{<\rho>}{M} \frac{d}{dM} F(>M) dM,
\end{displaymath}
where the function $F(>M)$ is the fraction of material in
objects of mass greater than $M$, and $<\rho>$ is the cosmic
mean density.  

\subsubsection{The Press-Schechter Ansatz}

	For gaussian theories, Press \& Schechter (1974) 
proposed the following:
\begin{displaymath}
F(>M) = \frac{1}{\sqrt{2\pi}} \int_{\nu_c}^\infty d\nu e^{-\nu^2/2},
\end{displaymath}
where $\nu_c\equiv \bar{\delta}/\sigma(M)$ is some critical, {\it 
linear} threshold for collapse.  Using this ansatz, derive
the Press-Schechter mass function (to within a factor of 2):
\begin{displaymath}
n(M) = \frac{1}{\sqrt{2\pi}}\frac{<\rho>}{M}\nu_c 
	\left(-\frac{dln\sigma}{dM}\right) e^{-\nu_c^2/2}.
\end{displaymath}

\subsubsection{Details}

	The above result undercounts the amount of mass in collapsed
objects.  To see this, show that
\begin{displaymath}
\int dM M n(M) \neq <\rho>.
\end{displaymath}
We must {\it arbitrarily} multiply by 2 to correct this expression
to find the Press--Schechter mass function.
This problem may be understood in two ways:  As pointed out by
Blanchard et al. (1992), this prescription only counts mass points
at the center of top--hat spheres satisfying the threshold; it does
not include particles within such spheres which are not at the
center.  Secondly, the prescription assumes that all collapsed
objects with a mass $>M$ satisfy the threshold criteria at the
smoothing scale corresponding to $M$.  Bond et al. (1991) 
show that this is in fact not the case (at least for
a sharp k--space filter); their formalism allowed them to
{\it derive} the factor of 2.

\subsection{The Temperature--Mass Relation}

	Here we will consider the origin of the temperature--mass
relation for virialized objects.  We start from the conservation
of energy and the virial theorem as applied to a thin shell of 
matter as it collapses onto the cluster:
\begin{eqnarray}
\label{cons}
K + U = const = U_{ta}\\
K + \frac{1}{2}U = 0,
\end{eqnarray}
where $K$ and $U$ represent, respectively, the kinetic energy and
gravitational potential energy of the shell, and $U_{ta}$ is its
potential (and hence total) energy at turn--around.  Each 
collapsing shell then contributes a kinetic energy of $K=-U_{ta}$
to the cluster, and the total kinetic energy may be written as
a sum over shells.  To proceed, assume that the collapse is 
spherical; there may be some reason to hope that this is not too 
far from reality for clusters, which often represent 
(in biased theories!) perturbations of several sigma 
(Bernardeau 1994).  

\subsubsection{Set-up}

	Show that an integral for $K_{tot}$ is
\begin{displaymath}
\label{Ktot}
K_{tot} = (4\pi<\rho>)^2\frac{G}{3} \int_0^{x_v} dx \frac{x^5}{\ata\xta},
\end{displaymath}
where $<\rho>$ is the mean {\it comoving} matter density, $x$ is 
a comoving {\it lagrangian} radius defined by $M(<x) = (4\pi/3)
x^3 <\rho>$, while $\xta$ is the comoving radius of the shell
at turn--around, when the expansion factor is $\ata$. 

\subsubsection{Specific Case}

	Consider the collapse of spherical density
perturbations with a power--law profile, $\dmean(x) 
\propto x^{-\gamma}$, in a critical universe ($\Omega_o=1$).
In this case,
\begin{displaymath}
\delta_{ta} = 1.06 = \dmean_o(x) \ata
\end{displaymath}
relates the radius of a shell to its turn--around epoch, $\ata$.
Show that the {\it mean, particle--weighted} temperature --
defined by $<T>_p \equiv (2/3)K_{tot}/(Nk)$, where $N$ is the 
total number of particles in the cluster and $k$ is Boltzmann's
constant -- of the cluster is 
\begin{displaymath}
<T>_p = (4\pi<\rho>)^2\frac{2G}{9}\frac{1}{\eta}
	\frac{\mu m_p}{k}\frac{(1+z)}{5-\alpha} M^{2/3}.
\end{displaymath}
The mean molecular weight of the particles is $\mu m_p$, and
$\eta \equiv \xta/x$.  Notice that the temperature has the 
form $M^{2/3} (1+z)$, and also that the constant of normalization 
in front depends on the density profile in the linear regime.

\subsubsection{Generalization}

	Prove, regardless of the density profile or the
cosmology, that $<T>_p$ always has the dependence $M^{2/3}
F(z)$, where $F(z)$ is an unknown function of the redshift only.
Once again, apply the spherical collapse model.

\subsubsection{Comments}

	The important point here is that, within the spherical 
collapse model, $<T>_p$ is always a function of $M^{2/3} F(z)$,
but that the coefficient in the relation depends on the cosmology
and the density profile.  This means that it may depend on the
spectrum of density perturbations, something which we ignored
during the lecture.  The other point is that $<T>_p$ is not
what is measured by the X-ray observations -- this is rather
the {\it emission weighted} temperature.  If the cluster
is not isothermal, they need not be the same.  On the other hand,
the SZ effect sees $<T>_p$.  This is one of the advantageous
aspects of the SZ modeling.

\subsection{The Sunyaev-Zel'dovich Effect}

	Inverse Compton scattering of CMB photons by a thermal
gas at temperature $T_e$ is described by the Kompaneets equation
(Kompaneets 1957) for the photon phase--space
distribution function, $n_\gamma$:
\begin{displaymath}
\label{komp}
\frac{\partial n_\gamma}{\partial y} = \frac{1}{x_e^2}
	\frac{\partial}{\partial x_e} \left(x_e^4\left(
	\frac{\partial n_\gamma}{\partial x_e} + n_\gamma + n_\gamma^2
	\right)\right)
\end{displaymath}
In this equation, $x_e\equiv h\nu/kT_e$, where $T_e$ is the 
temperature of the electron gas, 
and the {\it Compton-y parameter} is given by 
$dy\equiv (kT_e/m_ec^2)n_e\sigma_T cdt$.  Note that we distinguish
$x_e$ from $x\equiv h\nu/kT_o$, where $T_o=2.728$ is the
temperature today of the CMB, as used in the text. 

\subsubsection{General Solution}

	Show that the photon number density -- $N_\gamma = 
4\pi \int d\nu \nu^2 n_\gamma \propto \int dx_e x_e^2 n_\gamma$ --is 
conserved by this equation; what is the
physical reason for this?  Given this information, find the
most general {\it equilibrium} solution to the Kompaneets
equation (hint: it is an extension of a blackbody spectrum).

\subsubsection{First Order Solution}

	Galaxy clusters produce only a small perturbation to the
CMB spectrum, of order $y\sim 10^{-5}$.  Derive the first order
perturbative solution ($y<<1$) around a pure blackbody spectrum:
\begin{displaymath}
j_\nu(x) = 2\frac{(kT_o)^3}{(h_{\rm p}c)^2}\left(
	\frac{x^4e^x}{(e^x-1)^2}\left(\frac{x}{{\rm tanh}(x/2)}-4\right)\right).
\end{displaymath}
Notice that it is $x$ and not $x_e$ which is used in this expression.

\acknowledgments

	I would like to thank my collaborators on this work,
from whom I have learned much and with whom I have greatly enjoyed
working over the years: D. Barbosa, A. Blanchard, J. Oukbir, and
J. Silk.

\end{document}